\documentclass[aps,showpacs]{revtex4}
\usepackage{graphicx,psfrag}
\def\no{\noindent}
\def\bc{\begin{center}}
\def\ec{\end{center}}

\def\beq{\begin{equation}}
\def\eeq{\end{equation}}

\begin{document}

\title{Quantum transport in 3D Weyl semimetals: \\
Is there a metal-insulator transition?
}

\author{K. Ziegler}
\affiliation{
Institut f\"ur Physik, Universit\"at Augsburg,
D-86135 Augsburg, Germany}
\date{\today}

\begin{abstract}
We calculate the transport properties of three-dimensional Weyl fermions in a disordered
environment. The resulting conductivity depends only on the Fermi energy and the scattering rate.
First we study the conductivity at the spectral node for a fixed scattering rate and obtain 
a continuous transition from an insulator at weak disorder to a metal at stronger disorder.
Within the self-consistent Born approximation the scattering rate depends on the Fermi energy.
Then it is crucial that the limits of the conductivity for a vanishing Fermi energy and 
a vanishing scattering rate do not commute. As a result, there is also metallic behavior in 
the phase with vanishing scattering rate and only a quantum critical point remains as an 
insulating phase. The latter turns out to be a critical fixed point in terms of a 
renormalization-group flow.  
\end{abstract}

\maketitle

\section{Introduction}

Since the discovery of the fascinatingly robust transport properties of graphene 
\cite{novoselov05,zhang05,castroneto07b,abergel10},
there has been an increasing interest in other two-dimensional systems with similar spectral
properties, such as the surface of topological insulators \cite{kane05,koenig07,review_ti,qi11,schmeltzer13}. 
In all these systems the transport is dominated by a band structure, in which two bands 
touch each other at nodes. If the Fermi energy is exactly at or close to these nodes, the point-like
Fermi surface and interband scattering lead to particular transport properties, such as a robust 
minimal conductivity. Based on these results, an extension of the nodal spectral structure
to three-dimensional (3D) systems is of interest 
\cite{fradkin86a,fradkin86b,wan11,smith11,balents11,burkov11,xu11,kim12,young12,hosur12,wang12,
singh12,cho12,balents12,huse13,liu13,ruy13,kobayashi13,arovas13,herbut14,koshino14}.
In 3D the Fermi surface is a sphere with radius $|E_F|$ rather than the circular Fermi surface
in 2D, which is either occupied by electrons ($E_F>0$)
or by holes ($E_F<0$). For $E_F=0$ the conductivity vanishes in the absence of impurity scattering
in contrast to the minimal conductivity of the 2D system.
On the other hand, sufficiently strong impurity scattering leads to a conductivity at the node $E_F=0$.
Thus, an important difference between 2D and 3D Weyl fermions is that there exists a metal-insulator 
transition in the latter, which is driven by increasing disorder 
\cite{kobayashi13,herbut14,koshino14,roy14,brouwer14,gurarie14}.
This transition is similar to the metal-insulator transition caused by decreasing random gap fluctuations
in a system of 2D Dirac fermions \cite{ziegler09a,beenakker10}. On the other hand, it is quite different 
from an Anderson transition from a metallic state at weak scattering to an insulating state at strong scattering,
which is caused by Anderson localization at strong scattering \cite{anderson58,abrahams79}.

There is agreement between the various approaches, based on self-consistent, perturbative and numerical
methods, on the existence of a transition from a 3D Weyl semimetal at weak scattering to a diffusive metallic behavior
at stronger random scattering 
\cite{fradkin86a,fradkin86b,wan11,smith11,balents11,burkov11,xu11,kim12,young12,hosur12,wang12,
singh12,cho12,balents12,huse13,liu13,ruy13,kobayashi13,arovas13,herbut14,koshino14,roy14,brouwer14,gurarie14}. 
This transition can be characterized by a vanishing density
of states at the Weyl node (i.e., the scattering rate or the imaginary part of the self-energy) and a nonzero
density of states in the diffusive phase. However, the transport properties for the weak scattering regime
are still under discussion. In particular, a recent study
indicates that there is
a metal-metal transition rather than a insulator-metal transition for 3D Weyl fermions with a critical
point \cite{koshino14}. We will address this problem in the subsequent calculation, using a weak
scattering approach (WSA).

Calculations of quantum transport consist usually of two steps: Determining the scattering time 
(or scattering rate) 
within a self-consistent solution of the Dyson equation, also known as the self-consistent Born approximation
(SCBA), and determining the conductivity by a self-consistent solution of the Bethe-Salpeter equation (BSE).
This approach, in particular the solution of the BSE, is rather complex due to the existence of many modes.
Not all of them are relevant for the transport properties because some decay quickly. From this point of 
view it is easier to
project at the beginning only onto those modes which do not decay quickly but control the transport properties
on large scales. For a system with spectral nodes in a disordered environment these modes are a result of
a spontaneously broken chiral symmetry \cite{ziegler12c,ziegler97,ziegler98,ziegler12b}. We will employ this idea here to
3D Weyl fermions in order to calculate the conductivity.   
For this purpose it is important to identify the underlying symmetries of the two-particle Green's function.
Then spontaneous symmetry breaking is characterized by its non-vanishing order parameter which is the 
scattering rate. 
There is a metallic phase with long range correlations (i.e. diffusion), whereas in the insulating phase 
the symmetry remains unbroken.

The paper is organized as follows: In Sec. II we define the model and discuss the symmetry properties
of the two-particle Green's function. Then the DC conductivity is calculated within a weak scattering
approach (Sect. III), using an expansion in powers of the scattering rate $\eta$. This provides us
a formula for the DC conductivity, which is discussed in Sect. IV at the node (Sect. IVa) and away
from the node (Sect. IVb). Our discussion includes a comparison with the results of the Boltzmann
approach and with results from an approach based on the SCBA and BSE of Refs. \cite{ruy13,koshino14}.

\section{Model}

The three-dimensional Weyl Hamiltonian for electrons with momentum ${\vec p}$ is expanded 
in terms of Pauli matrices $\tau_j$ ($j=0,1,2,3$) as 
\beq
H=v_F{\vec\tau}\cdot{\vec p}-U\tau_0 \ \ \ {\rm with}\ \ {\vec\tau}=(\tau_1,\tau_2,\tau_3)
\label{ham00}
\eeq
with Fermi velocity $v_F$.
$U$ is a disorder term, represented by a random potential with mean $\langle U\rangle=E_F$ (Fermi energy)
and variance $g$. 
The average Hamiltonian $\langle H\rangle$ generates a spherical Fermi surface with radius $|E_F|$, and
with electrons (holes) for $E_F>0$ ($E_F<0$). Physical quantities are expressed in such units that $v_F\hbar=1$.

The electronic conductivity, obtained as the response to a weak external field with frequency $\omega\sim0$ 
\cite{thouless74,ziegler08a,abergel10}
\beq
\sigma(\omega)=-\frac{e^2}{2h}\omega^2\sum_{r}r_k^2 A_{r0}(\omega) 
\ ,
\label{conda1}
\eeq
is given by the correlation function of the Green's functions $(H\pm z)^{-1}$
\beq
A_{rr'}(\omega)=\lim_{\epsilon\to0}
\langle Tr_2[(H +z)^{-1}_{rr'}(H-z)^{-1}_{r'r}]\rangle
\ \ \ {\rm with}\ \ z=\omega/2+i\epsilon
\ ,
\label{corr1}
\eeq
where $\langle ... \rangle$ represents disorder average and $Tr_2$ is the trace with respect to the
Pauli matrix structure.
This expression, often called the two-particle Green's function, has two different energies $\pm z$
for the same Hamiltonian $H$ to create two independent Green's functions $(H\pm z)^{-1}$. 
Now we represent this two-particle Green's function by two different Hamiltonians and one energy:
We define the pair of Hamiltonians
\beq
H_\pm=p_1\tau_1+p_2\tau_2 \pm(p_3\tau_3-U\tau_0)
\ ,
\eeq
where $H_+=H$. The matrix transposition $^T$ relates $H_+$ and $H_-$ through the identity
\beq
\tau_1 H_\pm^T\tau_1 =-H_\mp
\ ,
\label{id1}
\eeq
since $p_j^T=-p_j$. 
This allows us to write for the correlation function (\ref{corr1})
\beq
A_{rr'}(\omega)=-\lim_{\epsilon\to0}
\langle Tr_2[(H_+ +z)^{-1}_{rr'}\tau_1(H_-^T+z)^{-1}_{r'r}\tau_1]\rangle
\ .
\label{corr1a}
\eeq
Instead of two different energies $\pm z$ and the same Hamiltonian $H$, the
two-particle Green's function has now the same energy $z$ but different Hamiltonians,
namely $H_+$ and $H_-^T$. The relation (\ref{id1}) and the representation (\ref{corr1a})
reveals an internal structure of the model which leads to the Hamiltonian
\beq
{\hat H}=\pmatrix{
H_+ & 0 & 0 & 0\cr
0 & H_- & 0 & 0\cr
0 & 0 & H_-^T & 0 \cr
0 & 0 & 0 & H_+^T \cr
} 
\ .
\label{8by8a}
\eeq
The Green's functions $(H_+ +z)^{-1}$ and $(H_-^T+z)^{-1}$ in Eq. (\ref{corr1a})
are just the first and the third diagonal element of the Green's function $({\hat H}+z)^{-1}$.
This indicates that the transport properties of the original Hamiltonian $H$, which requires
two different energies $\pm z$, are related to the transport properties of the extended Hamiltonian
${\hat H}$ at the same energy $z$.

The extended Hamiltonian ${\hat H}$, its symmetries and its relation to diffusive transport were
studied previously
\cite{ziegler12c,sinner12,ziegler97,ziegler12b}. In particular, it was found, together with property (\ref{id1}), 
that the matrix
\beq
{\hat S}=\pmatrix{
0 & 0 & \varphi_1 \tau_1 & 0 \cr
0 & 0 & 0 & \varphi_2 \tau_1 \cr
\varphi_1'\tau_1 & 0 & 0 & 0 \cr
0 & \varphi_2' \tau_1 & 0 & 0\cr
}
\label{symm_tr}
\eeq
with scalar variables $\varphi_j, \varphi_j'$ anticommutes with ${\hat H}$: ${\hat S}{\hat H}=-{\hat H}{\hat S}$. 
This relation implies a non-Abelian chiral symmetry \cite{ziegler97,ziegler12c}:
\beq
e^{\hat S}{\hat H}e^{\hat S}={\hat H}
\label{symmetry}
\eeq
which is a symmetry relation for the extended Hamiltonian ${\hat H}$ in Eq. (\ref{8by8a}). 
The term proportional to $z$ in the Green's function ${\hat G}(z)$ breaks this symmetry due to $e^{\hat S}\ne {\bf 1}$, and
therefore, $\lim_{z\to0}[{\hat G}(z)-{\hat G}(-z)]$ plays the role of an order parameter for
spontaneous symmetry breaking:
\beq
{\hat G}(z)-{\hat G}(-z)=-2z{\hat G}(z){\hat G}(-z)=-2z({\hat H}^2-z^2)^{-1}
\ .
\label{op}
\eeq
Since the diagonal elements of this expression are proportional to the density of states at the node when we take
the limit $z\to0$, a non-vanishing density of states indicates spontaneous symmetry breaking. The role of
a non-vanishing density of states at the node as an order parameter for a diffusive metallic phase was also
discussed in Refs. \cite{huse13,herbut14,roy14}.

Following the recipe of Ref. \cite{ziegler97}
the correlation function (\ref{corr1a}) can be expressed as a diffusion propagator.
This 
is used in the next section, where we focus on the long-range behavior
of $A_{rr'}(\omega)$ to calculate the conductivity $\sigma(\omega)$.

\section{Weak-scattering approach}

The scattering rate $\eta$ is defined by 
\beq
\eta=\frac{i}{2}\frac{Tr[\langle (H+i\epsilon )^{-1}\rangle - \langle (H-i\epsilon )^{-1}\rangle]}
{Tr[\langle (H+i\epsilon )^{-1}\rangle\langle (H-i\epsilon )^{-1}\rangle]}
\ .
\label{scatt_rate}
\eeq
This definition is motivated by the assumption of a complex self-energy for
the average one-particle Green's function $\langle (H+i\epsilon )^{-1}\rangle$, whose imaginary part is the 
scattering rate
(cf. App. \ref{sect:self-consist}). The corresponding scattering time $\tau$ is $\tau=\hbar/\eta$.
$\eta$ can either be calculated, for instance, within the SCBA \cite{abergel10,roy14,brouwer14} 
or it can be taken from experimental measurements. As discussed in the previous section, a non-vanishing
scattering rate indicates spontaneous symmetry breaking. Since the broken symmetry is continuous, there
exists a massless mode. The latter is reflected by the relation
\beq
\sum_{r'}A_{rr'}(\omega)=Tr_2\langle [(H+\omega/2 )^{-1} (H-\omega/2 )^{-1}]_{rr}\rangle =-\frac{1}{\omega}
Tr_2[\langle (H+\omega/2 )^{-1}_{rr}\rangle - \langle (H-\omega/2 )^{-1}_{rr}\rangle]
\label{relation1}
\eeq
which diverges for a vanishing symmetry breaking term $\omega\sim0$ due to long-range correlations. 
For the correlation function (\ref{corr1a}) a similar but more elaborate analysis yields a diffusion 
propagator \cite{ziegler09a}, whose Fourier components read
\beq
{\tilde A}_q(\omega)\approx -\frac{\eta}{g}\frac{1}{i\omega/2 +Dq^2}
\ .
\label{diff_prop}
\eeq
This agrees with (\ref{relation1}) for $q=0$ when we use the self-consistent approximation 
$\langle  (H+ i\epsilon+\omega/2 )^{-1}_{rr}\rangle\approx -\Sigma/g$ of App. \ref{sect:self-consist}.
This is not an accident but a consequence of the fact that the self-consistent approximation
represents the saddle point of the corresponding functional integral \cite{ziegler97}.
The prefactor of the $q^2$ term is the diffusion coefficient
\beq
D
=\frac{g\eta}{2}\int_k Tr_2\left(\frac{\partial (\langle H\rangle + i\eta)^{-1}}{\partial k_l}
\frac{\partial (\langle H\rangle -i\eta)^{-1}}{\partial k_l}\right)
\ .
\label{3ddiff}
\eeq
Thus, the DC limit $\omega\to0$ of the conductivity formula in (\ref{conda1}) and the correlation
function (\ref{diff_prop}) reproduce the Einstein relation
\beq
\sigma=\frac{e^2}{h}\frac{2\eta}{g} D
\ ,
\label{einstein_r}
\eeq
which gives with the right-hand side of Eq. (\ref{3ddiff}) for 3D Weyl fermions the integral
\beq
\sigma(\eta,E_F)=2\frac{e^2}{h}\eta^2\int_0^\lambda\frac{(\eta^2+k^2)^2+E_F^2(2\eta^2+2k^2/3+E_F^2)}
{[(\eta^2-E_F^2+k^2)^2+4\eta^2E_F^2]^2} \frac{k^2dk}{2\pi^2}
\label{cond2}
\eeq
with momentum cut-off $\lambda$. Thus, the conductivity depends on the disorder strength $g$ only through the
scattering rate $\eta$. 

A diffusion propagator can also be calculated from the BSE, as demonstrated recently for 3D Weyl fermions
\cite{ruy13,koshino14}. 
However, the derivation of the propagator (\ref{diff_prop})
from the symmetry (\ref{symmetry}) has the advantage that it is simpler and that we obtain the diffusion 
coefficient $D$ in (\ref{3ddiff}) directly as a quadratic form of Green's functions.

\section{Results and discussion}

In the following we present and discuss the results which are obtained from the conductivity 
$\sigma(\eta,E_F)$ in Eq. (\ref{cond2}). This expression is subtle in the limit of a vanishing 
scattering rate $\eta$, since the latter appears as $\eta^2$ in front of an integral that diverges 
for $\eta\to0$. This makes the conductivity very sensitive to the order of the
limits $E_F\to0$ and $\eta\to0$ in the case when the scattering rate vanishes at the node.
Since the conductivity depends on $\eta$ and $E_F$ separately,
we consider first properties exactly at the node $E_F=0$, where results are simple, and then 
the more complex results when $\eta$ depends on $E_F$. For the second part we employ the
SCBA to determine the function $\eta(E_F)$ and calculate the corresponding conductivity.

\subsection{Transport at the spectral node}

At the node ($E_F=0$) the DC conductivity in Eq. (\ref{cond2}) is reduced to the expression
\beq
\sigma =2\frac{e^2}{h}\eta^2\int_0^\lambda \frac{k^2}{(\eta^2+k^2)^2}\frac{dk}{2\pi^2}
=\frac{e^2\eta}{2\pi^2h}\left[\arctan(1/\zeta)-\frac{\zeta}{1+\zeta^2}\right] \ \ \ 
(\zeta=\eta/\lambda)
\ ,
\label{cond000}
\eeq
which becomes for $\lambda\gg\eta$
\beq
\sigma\sim \frac{ e^2}{4\pi h}\eta
\ .
\label{cond001}
\eeq
In contrast to the 2D case, where $\sigma=e^2/\pi h$, the 3D case gives a linearly increasing behavior with 
respect to the scattering rate. This result was derived
directly (i.e., without using Eq. (\ref{cond2})) by Fradkin some time ago \cite{fradkin86a}. 
With a disorder dependent scattering rate he also obtained a transition for a critical disorder strength $g_c$,
where the conductivity vanishes for $g\le g_c$ and increases linearly for $g>g_c$.

\begin{figure*}[t]
\includegraphics[width=9cm]{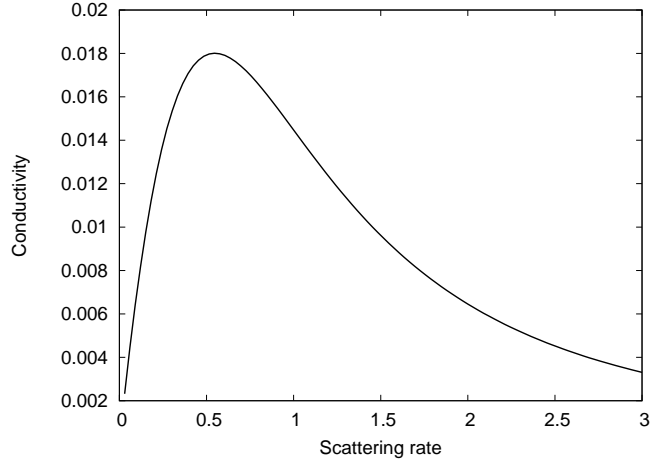}
\caption{
The conductivity of 3D Weyl fermions (\ref{cond2}) as a function of the scattering rate $\eta$ at Fermi 
energy $E_F=0$. It describes the crossover from the quantum regime for $\eta < 0.6$ to the classical 
Boltzmann regime for $\eta > 0.6$.
}
\label{fig:cond1}
\end{figure*}

The linear behavior indicates an unconventional transport because in the classical Boltzmann approach 
for one-band metals the conductivity {\it decreases} with increasing scattering rate: 
$\sigma_B=ne^2/m\hbar\eta$ ($n$: electron density, $m$: electron mass) \cite{ashcroft}.
This remains true when we include the band structure of the Weyl fermions in the Boltzmann approach:
$\sigma_B$ is nonzero at the node for any scattering rate and even diverges with vanishing disorder as \cite{koshino14}
\beq
\sigma_B=\frac{1}{2\pi}\frac{e^2 v_F^2 \hbar}{g}
\ ,
\label{cond_b}
\eeq
where 
$g$ is related to the density of impurities $n_i$ and the 
impurity potential $u_0$ by the equation $g=n_i u_0^2$ \cite{koshino14}. 
The disagreement between the expressions in (\ref{cond000}) and (\ref{cond_b})
can be explained by interband scattering, caused by particle-hole creation processes, which has been
ignored in the Boltzmann approach. On the other hand, the increasing behavior of (\ref{cond000}) 
 for small $\eta$ turns 
into a decreasing behavior for larger $\eta$, as one can see in Fig. \ref{fig:cond1}, indicating a 
crossover from quantum transport for weak scattering to conventional Boltzmann transport for stronger scattering.  

In this context it is also interesting to study the finite-size effects of 
the conductivity by considering a cubic system of finite length $L$.
The $\beta$--function $\beta=\partial \ln \sigma/\partial \ln L$ describes the finite-size scaling of the
conductivity. It can be calculated from Eq. (\ref{cond2}) by replacing the lower 
integration boundary with $1/L$, which assumes non-periodic boundary conditions.
Then $\sigma(\eta,L)$ in Eq. (\ref{cond000}) is a monotonically increasing function for increasing 
size $L$ with
\beq
\sigma(\eta,L)\sim \sigma^*-\frac{e^2}{h}\frac{\eta}{(L\eta)^2}
\eeq
for $L\eta\gg 1$, where $\sigma^*$ is the expression (\ref{cond001}). The corresponding $\beta$--function reads
\beq
\beta\sim2[1-\sigma(\eta,L)/\sigma^*]
\ ,
\label{betaf}
\eeq
which vanishes at the $\eta$-dependent fixed point $\sigma^*$.
This differs from the 2D case only by different fixed points $\sigma^*$, where in 2D it is a universal constant 
$\sigma^*=e^2/\pi h$ \cite{sinner14} and in 3D it is the $\eta$-dependent expression (\ref{cond001}).

\begin{figure*}[t]
\psfrag{beta}{$\beta=\frac{\partial \ln\sigma}{\partial\ln L}$}
\psfrag{sigma}{$\sigma$}
\includegraphics[width=6cm]{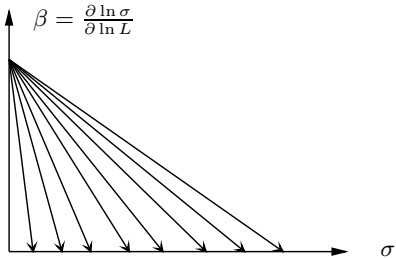}
\caption{
The $\beta$--function for different scattering rates, where the arrows indicate
the flow toward the fixed points. The $\beta$--function increases 
with $\eta$ and creates a line of fixed points $\sigma^*\sim\eta e^2/4h\pi$. 
This plot is based on an approximation near the fixed points according to Eq. (\ref{betaf}).
}
\label{fig:betaf}
\end{figure*}

It should be noticed that $\sigma^*$ is not a critical
point because it is an attractive fixed point. But since for 3D Weyl fermions $\sigma^*$ depends on the scattering
rate $\eta$, we have a line of fixed points for $\eta\ge 0$. Thus, the endpoint $\sigma^*=0$ for $\eta=0$
has the feature of a critical point because any change of $\eta$ drives us away from this endpoint,
as illustrated in Fig. \ref{fig:betaf}. 
It indicates a transition from an insulator ($\eta=0$) to a metal ($\eta>0$). The transition is driven by
increasing disorder, since the scattering rate is a monotonic function of the disorder strength
$g$. $\eta$ is also the order parameter for spontaneous symmetry breaking (\ref{op}), which can be calculated
from Eq. (\ref{scatt_rate}) within SCBA.
From the solution of the self-consistent equation (\ref{spe0}) we get for $\gamma=g/2\pi^2$ and $\eta\sim0$
the linear behavior
\beq
\eta\sim \frac{2\lambda}{\pi}\left(\lambda\gamma-1\right)\Theta(\lambda\gamma-1) \ \ \ (\gamma=g/2\pi^2)
\eeq
with the step function $\Theta$.
For $\gamma\le \gamma_c=1/\lambda$ we have no spontaneous symmetry breaking. Thus, $\eta$ as well as 
the DC conductivity vanish strictly. 
When we approach $\gamma_c$ from above there is linear behavior for the scattering rate, which agrees 
with the numerical calculation of Kobayashi et al. \cite{herbut14}. At the critical point itself we obtain
from the Einstein relation (\ref{einstein_r}) a finite diffusion coefficient
\[
D(g_c)\approx \frac{g_c e^2}{4\pi}
\ .
\]

The results of the DC conductivity from previous 
self-consistent studies, based on a combination of SCBA and a self-consistent solution of the 
BSE \cite{ruy13,koshino14}, are summarized and compared with our results of the WSA in Table \ref{table1}.
For sufficiently large scattering rates the Boltzmann approach, the solution of the BSE and the result of the WSA
agree reasonably well, reflecting a rather conventional transport. This indicates that quantum effects,
such as particle-hole pair creation, are dominated by impurity scattering. On the other hand, for smaller
values of the scattering rate the conductivity exhibits a larger variety of results: The Boltzmann
conductivity has a simple $1/g$ behavior, which is also found with the solution of the BSE in Ref. \cite{ruy13},
with a different constant prefactor though.
In contrast, the approximative analytic solution of the BSE in Ref. \cite{koshino14} has a characteristic dip down to
zero at a critical $g_c$ and increases for $g>g_c$ and for $g<g_c$:
\beq
\sigma=\sigma_1|1/g-1/g_c| , \ \ \ \sigma_1={\bar \sigma}\cases{
1 & for $g<g_c$ \cr
3 & for $g>g_c$ \cr
}
\ .
\label{scba_cond}
\eeq
When we compare this result with the WSA conductivity in Eq. (\ref{cond000}) it should be noticed that
the latter was obtained by sending $E_F\to0$ first and then $\eta\to0$. As mentioned at the
beginning of this Section the value of the conductivity depends on the way we take
these two limits. Although nothing has been said in Ref. \cite{koshino14} about the order
of the two limits to get (\ref{scba_cond}), we will study in Sect. \ref{sect:away}
the case when $E_F$ and $\eta$ go to zero simultaneously in Eq. (\ref{cond2}). Then we obtain
a result similar to (\ref{scba_cond}).

\begin{table}
\begin{center}
\begin{tabular}{|c|c|c|c|c|}
\hline 
{\bf at the node} & Boltzmann approach \cite{koshino14} & SCBA \& BSE \cite{ruy13} 
 & SCBA \& BSE \cite{koshino14} & SCBA \& WSA \\
\hline 
scattering rate  $\eta$ & $0$ & 0 & $2(g/g_c-1)/\pi$ & $2(g/g_c-1)/\pi$ \\ 
\hline 
conductivity $\sigma$ & $e^2 v_F^2 \hbar/2\pi g$  & $4e^2v_F^2/g h$ & $\sigma_1 e^2/4\pi h$ & Eq. (\ref{cond000}), Fig.
\ref{fig:cond1} \\
\hline
{\bf away from the node $g<g_c$} & & & & \\ 
\hline
scattering rate $\eta$ & $\propto E_F^2$ & $\hbar g E_F^2/8\pi v_F^3$ & $E_F g_c/(g_c-g)$ & Eq. (\ref{scattering1}) \\ 
\hline 
conductivity $\sigma$ & $\sigma_B(0)(1+6E_F^2/E_0^2)$ & $4e^2v_F^2/g h$ & $\propto E_F g_c/(g_c-g)$ & Fig. \ref{fig:cond3} \\
\hline 
\end{tabular}
\caption{The scattering rate $\eta$ and the conductivity at the node calculated with
three different methods. The translation from Ref. \cite{koshino14}
is $g=n_i u_0^2$ and the SCBA coefficient $\sigma_1$ is given in Eq. (\ref{scba_cond}). 
}
\label{table1}
\end{center}
\end{table}

\subsection{Transport away from the spectral node}
\label{sect:away}

\begin{figure*}[t]
\includegraphics[width=9cm]{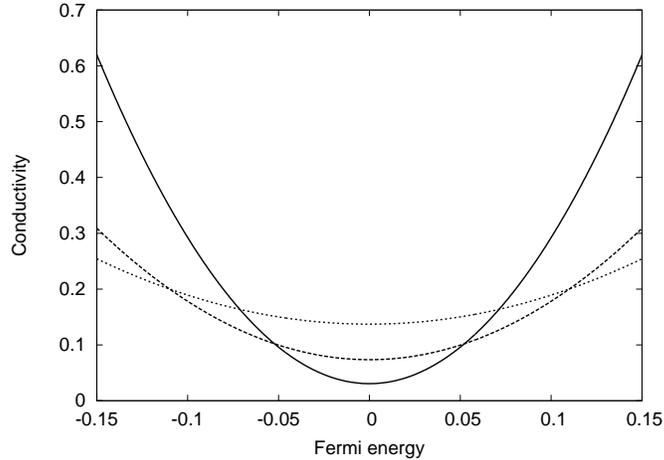}
\caption{
The conductivity (\ref{cond2}) as a function of the Fermi energy $E_F$ at fixed scattering rates 
$\eta=0.02$ (full curve), $\eta=0.05$ (dashed curve) and $\eta=0.1$ (dotted curve).
}
\label{fig:cond2}
\end{figure*}

The conductivity as a function of the Fermi energy is plotted at fixed scattering rates 
in Fig. \ref{fig:cond2}. As we increase the scattering rate the effect of the node is washed
out and the conductivity becomes flatter. This is similar to the behavior in Fig. \ref{fig:cond1}.
In other words, impurity scattering supports transport near the node whereas it suppresses it
further away.
Thus, we can distinguish a regime close to the node, where the conductivity increases
with the scattering rate, and a more conventional regime further away from the node, where the 
conductivity decreases with the scattering rate, as also described by the Boltzmann approach. 

\begin{figure*}[t]
\psfrag{E_F}{$E_F$}
\psfrag{disorder}{$\Delta$}
\includegraphics[width=9cm]{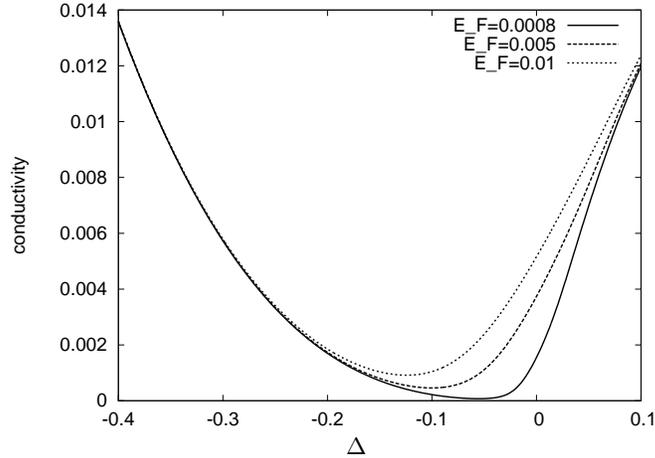}
\caption{
The conductivity (\ref{cond2}) as a function of the disorder parameter $\Delta=\gamma\lambda-1$ with 
a self-consistent scattering rate 
$\eta(\Delta)$ for different values the Fermi energy $E_F$. The curves tend to reach the critical
point $\Delta=0$ by a sharp cusp as the Fermi energy approaches the node $E_F=0$. Further
away from the critical point the conductivity is much less sensitive to a change of $E_F$.
}
\label{fig:cond3}
\end{figure*}

So far we have considered the case that $\eta$ and $E_F$ are independent. However, in general 
the scattering rate depends on $\gamma$ and $E_F$. For instance, the self-consistent calculation in
App. \ref{sect:self-consist}, based on the SCBA or saddle-point approximation, creates a 
scattering rate in Eq. (\ref{scatt3}) that depends on the Fermi energy:
\beq
\eta=Re\left[\frac{(\gamma\lambda-1)}{\gamma\pi} 
+\sqrt{\frac{(\gamma\lambda-1)^2}{\gamma^2\pi^2}-\frac{2iE_F}{\gamma\pi}}
\right]
\ .
\label{scattering1}
\eeq
The behavior of the conductivity in (\ref{cond2}) is affected by this result, since the 
limits $\eta\to0$ and $E_F\to0$ are not independent anymore.
(\ref{scattering1}) has two typical regimes, namely $\eta\propto E_F$ near the critical point $\gamma\lambda=1$,
which leads to $\sigma\propto E_F$, and $\eta\propto E_F^2$ for $\gamma\lambda\ll 1$, which
leads to a non-vanishing conductivity for $E_F\to0$. This implies that the conductivity
vanishes only for $\gamma\lambda\sim 1$, whereas it nonzero above and below $\gamma\lambda\sim 1$.
Thus, there is no insulating phase but only an insulating point for $\gamma\lambda=1$, in agreement
with the analytic result of Ref. \cite{koshino14}. The behavior of the conductivity as a function
of $\gamma\lambda -1$ is plotted for different values of the Fermi energy in Fig. \ref{fig:cond3}.
It should be noticed, though, that the transport behavior is determined by the $E_F$ dependence of the
scattering rate of the special form in Eq. (\ref{scattering1}). Using another form of the scattering rate as
a function of the Fermi energy can lead to a substantially different behavior of the conductivity
near the node. An example was observed in Ref. \cite{koshino14} within a numerical solution
of the SBCA and the BSE, where the scattering rate is exponentially small for $\gamma<\gamma_c$. In this
case a vanishing conductivity was found at the node also for $\gamma<\gamma_c$.

\section{Conclusions}

We have studied the DC conductivity of 3D Weyl fermions in the presence of random
scattering. The relevant parameters in the conductivity (\ref{cond2}) are the scattering rate $\eta$, which is an 
order parameter for spontaneous chiral symmetry breaking, and the Fermi energy $E_F$. Exactly at the node 
$E_F=0$ there is a metal-insulator transition with a diffusive metal for $\eta>0$ and an insulator for $\eta=0$.
The conductivity is linearly increasing with $\eta$ up to a maximal value and decreases for stronger
scattering rates, as illustrated in Fig. \ref{fig:cond1}. This non-monotonic behavior is in contrast 
to the constant conductivity in the corresponding 2D system. It reflects the fact that the increased phase
space of the 3D Weyl fermions suppresses the conductivity for weak scattering but also that stronger scattering
implies a screening of the node such that the Boltzmann approach eventually becomes applicable. 
Further away from the node the behavior depicted in Fig. \ref{fig:cond2} agrees qualitatively with that
of the 2D system \cite{ziegler12b}, which was also obtained in a quasiclassical approach \cite{trushin08}.
The latter diverges as one approaches the node, which indicates that the full quantum approach
is necessary near the node. 

The critical behavior at the node describes an unconventional phase transition for $\gamma=\gamma_c$
which is driven by quantum fluctuations: In contrast to a conventional transition the symmetry
broken phase with $\eta>0$ is characterized by robust diffusion whereas the phase with unbroken 
symmetry ($\eta=0$) has a subtle behavior in terms of the conductivity because it is very sensitive
to the limit $E_F\to0$. Thus, it is possible that we either have an insulating phase with vanishing
conductivity when the scattering rate vanishes slowly with $E_F$ or a metallic phase when the scattering
rate vanishes sufficiently fast with $E_F$. In the case of an SCBA calculation for $\eta$ there is only a 
quantum critical point in the transport properties and metallic behavior above and below this critical
point. It cannot be ruled out, though, that a different calculation of $\eta$ leads to a different behavior. 
Thus, our discussion of the delicate limits $\eta,E_F\to0$ clarifies some of the contradicting results in the 
literature about the presence of a metal-insulator transition for 3D Weyl fermions \cite{ruy13,koshino14}.
Moreover, the fact that the conductivity depends only on $\eta$ and $E_F$ allows us to determine $\eta$
independently with other approximations than the SCBA, and to insert the result into the conductivity 
(\ref{cond2}). A possible step in this direction is a correction to the SCBA \cite{brouwer14} or 
perturbative renormalization-group approach in $d-2$--expansion. The latter gives $\eta(E)\propto E^{1.3}$
\cite{herbut14,roy14}. This result would lead to a vanishing conductivity below the critical point.

In the regime $0<\gamma<\gamma_c$ disorder may also affect physical properties of Weyl fermions in another way. 
The reason is the existence of non-uniform solutions of the SCBA with an exponentially small 
contributions to $\eta$, similar to Lifshitz tails in the density of states of disordered systems \cite{pastur}.
In the case of 2D Weyl fermions this has been discussed in Ref. \cite{simon00}.
The problem, however, is always that the self-consistent equation is nonlinear and has many non-uniform
solutions. Under certain plausible assumptions for the solutions $\eta$, the existence of exponentially 
small contributions for $\gamma<\gamma_c$ has been discussed for 3D Weyl fermions in Ref. \cite{huse13}.
The corresponding states might be localized then with no contribution to the conductivity at $T=0$.
However, for $T>0$ thermally activated electrons may hop between patches of localized states and    
provide a hopping conductivity \cite{mott}. Whether or not resonant tunneling without spontaneous symmetry 
breaking can occur in this case is an open question. 

\vskip0.5cm

\no
Acknowledgement: I am grateful to David Schmeltzer for an extended discussion of Weyl fermions.

\appendix

\section{Self-consistent approximation}
\label{sect:self-consist}

The first step is to study spontaneous symmetry breaking of the symmetry (\ref{symmetry}) by a non-zero
scattering rate within SCBA, following a similar approach as given in Ref. \cite{koshino14}.
The average one-particle Green's function then reads
\beq
\langle (H+z)^{-1}\rangle\approx (\langle H\rangle + z+\Sigma)^{-1}
\ ,
\label{scba}
\eeq
and the self-energy is given by
\beq
\Sigma=-g(\langle H\rangle+\Sigma+i\epsilon)^{-1} 
\ .
\eeq
The imaginary part of the self-energy $\Sigma$ is a scattering rate $\eta$.
Then the self-energy reads in our case with the momentum cut-off $\lambda$
\beq
\Sigma=\gamma\alpha\left[\lambda-\frac{\alpha }{2}
\log\left(\frac{\alpha +\lambda}{\alpha -\lambda}\right)\right] 
\ \ \ (\gamma=g/2\pi^2,\ \ \alpha =E_F +\Sigma) 
\ .
\label{scba1}
\eeq
For small $E_F$ near the node we expand Eq. (\ref{scba1}) in powers of $\alpha$ up to second order to obtain
\beq
\Sigma\sim -E_F+\frac{i(\gamma\lambda-1)}{\gamma\pi} + 
i\sqrt{\frac{(\gamma\lambda-1)^2}{\gamma^2\pi^2}-\frac{2iE_F}{\gamma\pi}}
\ .
\label{scatt3}
\eeq
The real part of $\Sigma$ provides a shift of the Fermi energy:
\beq
\Sigma\sim -E_F + i\sqrt{-\frac{2iE_F}{\gamma\pi}}
=-E_F + e^{i\pi/4}\sqrt{\frac{2E_F}{\gamma\pi}}
\ ,
\eeq
where the sign is chosen such that we have a positive scattering rate.

At the node $E_F=0$ the self-consistent eq. (\ref{scba1}) reduces to $\eta=\eta I$ with
\[
I=\gamma\left[
\lambda-\eta\arctan(\lambda/\eta)\right] 
\ .
\]
There are two solutions, namely $\eta=0$ and $\eta\ne0$ with
\beq
\lambda\gamma=\frac{1}{1-\zeta\arctan(1/\zeta)} , \ \ \ \ \zeta=\eta/\lambda
\ .
\label{spe0}
\eeq
A nonzero $\eta$ reflects spontaneous symmetry breaking with respect to
(\ref{symmetry}). Such a solution exists for (\ref{spe0}) only at
sufficiently large $\gamma$. Moreover, $\eta$ vanishes continuously as we reduce $\gamma$. 
A nonzero $\eta$ is proportional to the density of states at the Fermi
level. However, even for $\eta=0$ there can be a nonzero local density of states due to localized energy levels,
which are not counted in $\eta$ within the SCBA.
For $\zeta\sim 0$ we obtain the linear behavior
\[
\zeta\sim\frac{2}{\pi}(\gamma\lambda -1)
\ .
\]

\end{document}